\documentstyle[11pt,newpasp,twoside,epsf]{article}
\def\emphasize#1{{\sl#1\/}}

\newfont{\rmsmall}{cmr10 scaled 900}

\def\edcomment#1{\iffalse\marginpar{\raggedright\sl#1\/}\else\relax\fi}
\reversemarginpar

\begin{document}
\title{The AST/RO Survey of the Galactic Center Region}
 \author{Antony A. Stark}
\affil{Smithsonian Astrophysical Observatory, 60 Garden St. MS 12,
Cambridge, MA 02138 USA}

\begin{abstract}
AST/RO is a 1.7m diameter submillimeter-wave telescope 
at the geographic South Pole.
A key AST/RO project is the mapping
of C\,{\rmsmall I} and CO $J=4\rightarrow3$
and $J=7\rightarrow6$ \, emission from the inner 
Milky Way (Martin et al. 2003).
These data are released for general use.
\end{abstract}

The Antarctic Submillimeter Telescope and Remote Observatory (AST/RO)
is a 1.7 m diameter single-dish instrument which has been
observing in the submillimeter-wave atmospheric windows
for eight years
(Stark et al. 2001, Stark 2003).
Essential to AST/RO's capabilities is its location at
Amundsen-Scott South Pole Station, an exceptionally cold,
dry site
which has unique logistical opportunities and challenges.
Observing time on AST/RO is available on a proposal basis.

The distribution of
molecular gas in the Galaxy is known from extensive
and on-going surveys in CO and $^{13}$CO $J=1\rightarrow0$ and
$J=2\rightarrow1$; these are
spectral lines which trace molecular gas.
These lines alone do not, however, determine the
excitation temperature, density, or cooling
rate of that gas.
Observations of C\,{\rmsmall I}  and the mid-$J$ lines of CO and $^{13}$CO
provide the missing information, showing a more complete picture
of the thermodynamic state of the molecular gas, highlighting
the active regions, and looking into the dense cores.
AST/RO can measure
the dominant cooling lines of molecular material in
the interstellar medium:
the ${ {}^{\rm 3}\! P_{\rm 1}\rightarrow{}^{\rm 3}\! P_{\rm 0}}$
(492 GHz) and 
${ {}^{\rm 3}\! P_{\rm 2}\rightarrow{}^{\rm 3}\! P_{\rm 1}}$
(809 GHz)
fine-structure lines of atomic carbon (C\,{\rmsmall I})
and the $J=4\rightarrow3$ (461 GHz) and $J=7\rightarrow6$ (807 GHz) rotational lines 
of carbon monoxide (CO).
These measurements can then be modelled
using the large velocity gradient
(LVG) approximation, and the gas temperature and density thereby determined.
Since the low-$J$ states of CO are in local thermodynamic
equilibrium (LTE) in almost all molecular gas,
measurements of mid-$J$ states are critical to
achieving a model solution
of the radiative transfer
by breaking the degeneracy between beam filling factor
and excitation temperature.

Among the key AST/RO projects is mapping of 
the Galactic Center Region.
Sky coverage as of 2002 is
$-1 \fdg 3 < \ell < 2 \deg$,
$-0 \, \fdg 3 < b < 0 \, \fdg 2$ with
$0 \farcm 5$ spacing, resulting in spectra of 3 transitions at
24,000 positions on the sky.  
Kim et al. (2002) and Martin et al. (2003) 
describe the data, which
are available 
on the AST/RO website\footnote{\tt http://cfa-www.harvard.edu/ASTRO}
for general use.
\begin{figure}[t!]
\plotone{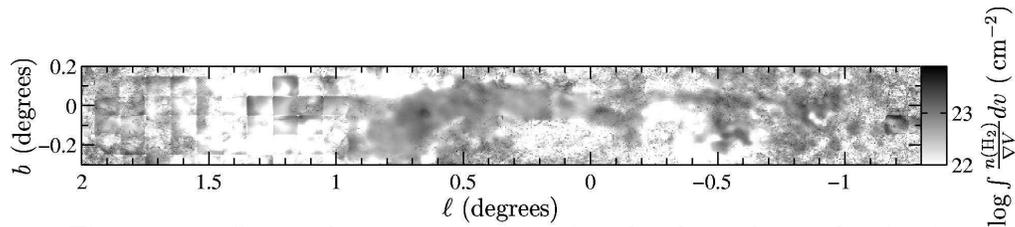}
\caption{Greyscale representation of molecular column density in the Galactic Center
Region, from an LVG model using AST/RO survey data (Martin et al. 2003).}
\end{figure}
The $\mathrm{C}\,\scriptstyle{\rm I}$ emission has a spatial extent 
similar to the low-$J$ CO emission, but is more diffuse. 
The CO $J = 4 \rightarrow 3$
emission is also found to be essentially coextensive with
lower-$J$ transitions of CO, indicating that even the $J=4$ state is
in LTE most places; in contrast, the CO $J = 7 \rightarrow 6$ emission
is spatially confined to far smaller regions.
Applying an LVG model to these data together with
data from the Bell Labs 7-m (Bally et al. 1988)
yields maps
of gas density and temperature as a
function of position and velocity for the entire region.
Kinetic temperature is found to decrease from
relatively high values ($>70$ K) at cloud edges to lower values ($<50$ K)
in the interiors.
Typical pressures in the Galactic Center gas are
$n(\mathrm{H_2}) \cdot T_{kinetic} \sim 10^{5.2} \, \mathrm{K \, cm^{-3}}$.

Above is a map of molecular hydrogen column density.
It is often assumed that molecular hydrogen column
density is proportional to the brightness of the
$J = 1 \rightarrow 0$ CO line.
The column densities estimated
using AST/RO data deviate in places
{\emphasize{by two orders of magnitude}} from this simple assumption.
These discrepancies are caused by
variations in excitation and optical depth.

Galactic Center gas that Binney et al. (1991) 
identify as being on $x_2$ orbits has a density 
near $10^{3.5} \, \mathrm{cm ^{-3}}$,
which renders 
it only marginally stable
against gravitational coagulation into one or two giant clouds (Elmegreen 1994).
This suggests a relaxation oscillator mechanism for starbursts, where
inflowing gas accumulates in a ring at 300 pc radius for approximately 400 million
years, until the critical density is reached, and the resulting 
instability leads to the sudden deposition 
of $10^7 {\mathrm{M_{\sun}}}$ of gas onto the Galactic Center.

\acknowledgments
Support was provided by NSF grant OPP-0126090.

\end{document}